\documentclass[preprint,nofootinbib]{revtex4-1} 
\usepackage{amsmath}  % needed for \tfrac, \bmatrix, etc.
\usepackage{amsfonts} % needed for bold Greek, Fraktur, and blackboard bold
\usepackage{graphicx} % needed for figures
\usepackage{color} 
\usepackage{amssymb}
\usepackage{appendix}
\usepackage{enumerate}

\begin{document}

\title{Comment on “Massive electrodynamics and the magnetic monopoles”}

\author{Michael Dunia}
\email{michaelrobertdunia@mail.fresnostate.edu}
\affiliation{Department of Physics, California State University Fresno, Fresno, CA 93740-8031, USA}
\author{Timothy J. Evans}
\email{tevans559@mail.fresnostate.edu}
\affiliation{Department of Physics, California State University Fresno, Fresno, CA 93740-8031, USA}
\author{Douglas Singleton}
\email{dougs@mail.fresnostate.edu}
\affiliation{Department of Physics, California State University Fresno, Fresno, CA 93740-8031, USA}

\date{\today}

\begin{abstract}
In this paper we correct previous work on magnetic charge plus a photon mass. We show that contrary to previous claims this system has a very simple, closed form solution which is the Dirac string potential multiplied by a exponential decaying part. Interesting features of this solution are discussed namely: (i) the Dirac string becomes a real feature of the solution; (ii) the breaking of gauge symmetry via the photon mass leads to a breaking of the rotational symmetry of the monopole's magnetic field; (iii) the Dirac quantization condition is potentially altered.               
\end{abstract}

\maketitle
 
\section{Maxwell's equations with magnetic charge and photon mass}
 
Two common extensions of electrodynamics are to add a magnetic charge or a photon mass into Maxwell's equations. The former extension was consider by Dirac \cite{dirac,dirac1} through the introduction of a singular 3-vector potential, which led to a quantization conditions between electric and magnetic charge. The latter extension was proposed by Proca \cite{proca}.  Reference \cite{joshi} considers both extensions together. The aim of this work is to correct the analysis presented in \cite{joshi} and remark on some of the unusual features that arise from the simple solution to this system that we find here.

Maxwell's equations with both magnetic charge and photon mass can be written in 4-vector forms as \cite{joshi}
\begin{eqnarray}
\label{maxwell4-source}
&& \partial _\nu F^{\nu \mu} + m ^2 A^\mu = 4 \pi J^\mu _{(e)} \\
\label{maxwell4-source-m}
&& \partial _\nu {\cal F}^{\nu \mu} = 4 \pi J^\mu _{(m)} ~,
\end{eqnarray}   
 where $J^\mu _{(e)} = ( \rho_e, {\bf J}_e )$ and $J^\mu _{(m)} = ( \rho_m, {\bf J}_m )$ are the electric and magnetic 4-vector currents \footnote{We use Gaussian units and set $c=1$. Since we are using Gaussian units, we mostly use the $2^{nd}$ edition of reference \cite {jackson} since this also uses Gaussian units}. The field strength tensor and its dual are defined via the 4-vector potential $A^\mu = (\phi, {\bf A})$ as $F^{\mu \nu} = \partial ^\mu A^\nu - \partial^\nu A^\mu$ and ${\cal F}^{\mu \nu} = \frac{1}{2} \epsilon^{\mu \nu \alpha \beta} F_{\alpha \beta}$. Finally, $m ^2 A^\mu$ is the Proca photon mass term, with $m$ being the photon mass. In general, due to symmetry, the divergence of the dual field strength tensor is zero $\partial _\nu {\cal F}^{\nu \mu}=0$. Below we will discuss in what sense one can have a magnetic charge source term in the right hand side of \eqref{maxwell4-source-m}.

Following reference \cite{joshi} we carry out our analysis of this system in 3-vector form. Writing out the 4-vector form of Maxwell's equations from \eqref{maxwell4-source} and \eqref{maxwell4-source-m} in 3-vector notation we have

\begin{eqnarray}
\label{e-maxwell3-source}
&&\nabla \cdot {\bf E} + m ^2 \phi = 4 \pi \rho _{e}  \\
\label{b-maxwell3-source-1}
&& \nabla \times {\bf B} - \frac{\partial {\bf E}}{\partial t} + m ^2 {\bf A} = 4 \pi {\bf J}_{e}  \\ 
\label{b-maxwell3-source}
&& \nabla \cdot {\bf B} = 4 \pi \rho _{m}  \\
\label{e-maxwell3-source-1}
&& \nabla \times {\bf E} + \frac{\partial {\bf B}}{\partial t} = - 4 \pi {\bf J}_{m}   ~.
\end{eqnarray}

We start with the system of a  magnetic charge, $g$, in the presence of a non-zero photon mass, $m$. The magnetic charge density is $\rho _{m} = g \delta ^3 ({\bf r})$. We first assume there is no electric charge (later we will consider both electric and magnetic charge together) we have $\rho _{e} = 0$ and ${\bf J}_{e} = 0$. We fix the magnetic charge to be at rest so ${\bf J}_{m} = 0$. For this set-up the fields are time-independent, $\partial _t {\bf E} = \partial _t {\bf B} = 0$. Since $\rho _{e} = 0$ the scalar potential is zero, $\phi =0$, which implies a zero electric field ${\bf E}=0$. Using all these  conditions the equations  \eqref{e-maxwell3-source} \eqref{b-maxwell3-source-1} \eqref{b-maxwell3-source} \eqref{e-maxwell3-source-1} reduce to
\begin{eqnarray}
\label{max-mono-m}
&& \nabla \times {\bf B} + m^2 {\bf A} = 0  \\ 
\label{max-mono-m1}
&& \nabla \cdot {\bf B} = 4 \pi \rho _{m}  = 4 \pi g \delta ^3 ({\bf r}) ~.
\end{eqnarray}
We first review the solution to \eqref{max-mono-m} and \eqref{max-mono-m1} for the massless case when $m=0$. In this case equation \eqref{max-mono-m1} is solved by the Dirac string potentials ${\bf A}^{(0)} _\pm (r, \theta) = \frac{g}{r} \left( \frac{\pm 1 - \cos \theta }{\sin \theta} \right) {\bf {\hat \varphi}}$. The unit vector in the $\varphi$-direction can be expanded as ${\bf {\hat \varphi}} = \frac{1}{\sin \theta} {\bf {\hat z}} \times {\bf {\hat r}}$ with ${\bf {\hat r}} =\frac{{\bf r}}{r}$ being the unit vector in the radial direction. This vector potential is valid everywhere except at $r=0$ and along $\theta=\pi$ or $\theta=0$ for the $+$ and $-$ signs respectively. It is also possible (and useful) to write the Dirac string potential in cylindrical coordinates as ${\bf A}^{(0)} _\pm (\rho, z) = \frac{g}{\rho} \left( \pm 1 - \frac{z}{\sqrt{\rho^2 + z^2}} \right) {\bf {\hat \varphi}}$ where the relationship of the $\rho$ and $z$ cylindrical coordinates to the spherical polar coordinates is given via $\rho = r \sin \theta$ and $z = r \cos \theta$

Recent work \cite{shnir} \cite{adorno} \cite{heras} has emphasized the presence of an explicit string piece to the magnetic field, in addition to the Coulomb part of the magnetic field. In reference \cite{heras} a regularized version of the of the Dirac string potential in cylindrical coordinates was considered
\begin{equation}
\label{regular-B}
{\bf A}^{(0)}_{\pm~\epsilon} (\rho, z) =  \frac{g\Theta (\rho -\epsilon)}{\rho}  \left( \pm 1 - \frac{z}{\sqrt{\rho^2 + z^2 + \epsilon ^2}} \right) {\bf {\hat \varphi}} ~.
\end{equation}
In \eqref{regular-B} $\Theta (x)$ is the step function which equals $1$ when the argument is positive and equals $0$ when the argument is negative, and $\epsilon$ is an infinitesimal quantity which is taken to zero at the end. Taking the curl of \eqref{regular-B} and taking the limit $\epsilon \to 0$ at the end yields  \cite{heras}
\begin{equation}
\label{regular-B2}
{\bf B}^{(0)} = \lim _{\epsilon \to 0}\nabla \times {\bf A}^{(0)}_{\pm~\epsilon} = g \frac{{\bf \hat r}}{r^2} \pm 4 \pi g \delta (x) \delta (y) \Theta (\mp z) {\bf {\hat z}}
\end{equation}
Usually, only the first, Coulombic term, $g \frac{{\bf \hat r}}{r^2}$ is written down, but a careful analysis (see \cite{shnir} \cite{adorno} and also \cite{heras} for a recent, pedagogical and thorough exposition) shows the existence of the second, string term, which is required to make sure that the divergence of a curl is zero.  In detail one can see that $\nabla \cdot \left( g \frac{{\bf {\hat r}}}{r^2} \right) = 4 \pi g \delta ^3 ({\bf r})$ and also that $\nabla \cdot \left(\pm 4 \pi g \delta (x) \delta (y) \Theta (\mp z) {\bf {\hat z}} \right) = -4 \pi g \delta ^3 ({\bf r})$, so that in total $\nabla \cdot {\bf B}^{(0)} =  \lim _{\epsilon \to 0} \nabla \cdot (\nabla \times {\bf A}^{(0)}_{\pm~\epsilon}) = 0$. Thus it is only the first term in ${\bf B}^{(0)}$ ({\it i.e.} $g \frac{{\bf {\hat r}}}{r^2}$) that gives the $\delta$-function magnetic point source. The second term in ${\bf B}^{(0)}$ ({\it i.e.} $\pm 4 \pi g \delta (x) \delta (y) \Theta (\mp z) {\bf {\hat z}}$) does not have a zero curl but rather gives $\nabla \times {\bf B}^{(0)}  = \pm 4 \pi g \Theta (\mp z) \left[ \delta (x) \delta ' (y) {\bf \hat x} - \delta ' (x) \delta (y) {\bf \hat y} \right]$, where the primes indicate differentiation with respect to the argument, $x$ or $y$. Thus $\nabla \times {\bf B}^{(0)} \ne 0$ in apparent violation of \eqref{max-mono-m} with $m=0$, but this non-zero curl for ${\bf B}$ represents an effective current density associated with the Dirac string. By imposing the Dirac quantization condition of $qg= n \frac{\hbar}{2}$ the string is made ``invisible" for the most part. Usually the effect of the string on the divergence and curl of ${\bf B}^{(0)}$ is not discussed in detail. The recent works \cite{adorno} \cite{heras} have pointed out these subtle issues connected with the Dirac string. 

We now move on to the massive case when $m \ne 0$ in \eqref{max-mono-m}. Reference \cite{joshi} gave a complex, not analytic form for the solution to equations \eqref{max-mono-m} and \eqref{max-mono-m1}. Here we show that the proposed solution of \cite{joshi} is not correct; that in fact these equations have a simple, closed form solution which is simply the Dirac string potential multiplied by a Yukawa factor $e^{-mr}$. This mirrors the Yukawa potential for electric charge which is just a Coulomb scalar potential multiplied by $e^{-mr}$.

Our ``guess" at a solution to equations \eqref{max-mono-m} and \eqref{max-mono-m1} is to take ${\bf A}^{(0)} _\pm$ and multiply it by $e^{-mr}$~  giving \cite{evans}
\begin{equation}
\label{b-yukawa}
{\bf A}_\pm = e^{-mr} {\bf A}^{(0)} _\pm (r) = g \frac{e^{-m r}}{r} \left( \frac{\pm 1 - \cos \theta }{\sin \theta} \right) {\bf {\hat \varphi}} \\
\end{equation}
To obtain the magnetic field from \eqref{b-yukawa} requires that we take the curl of the vector potential in \eqref{b-yukawa}. This takes some care due to the singularity along the $\pm z$-axis. Following appendix D of \cite{heras} we convert the vector potential in \eqref{b-yukawa} into cylindrical coordinates and regularize it via an infinitesimal $\epsilon$ to give
\begin{equation}
    \label{A-regular}
    {\bf A}_{\pm ~ \epsilon} = \frac{g \Theta (\rho - \epsilon)}{\rho} \left( \pm 1 - \frac{z}{\sqrt{\rho^2 + z^2 + \epsilon^2}} \right) e^{-m\sqrt{\rho^2 + z^2 + \epsilon ^2}} {\bf {\hat \varphi}} ~. 
\end{equation}
In the end we will take $\epsilon \to 0$. 
The magnetic field coming from \eqref{A-regular} is ${\bf B}_\pm = \nabla \times  ({\bf A} _{\pm ~ \epsilon} (r) )$ which yields
\begin{eqnarray}
    \label{b-yukawa-a}
    {\bf B} _\pm &=&  e^{-m\sqrt{\rho^2 + z^2 + \epsilon ^2}} \nabla \times \left[ \frac{g \Theta (\rho - \epsilon)}{\rho} \left( \pm 1 - \frac{z}{\sqrt{\rho^2 + z^2 + \epsilon^2}} \right) {\bf {\hat \varphi}} \right] \nonumber \\
    &+&  \nabla \left( e^{-m\sqrt{\rho^2 + z^2 + \epsilon ^2}} \right) \times \left[ \frac{g \Theta (\rho - \epsilon)}{\rho} \left( \pm 1 - \frac{z}{\sqrt{\rho^2 + z^2 + \epsilon^2}} \right) {\bf {\hat \varphi}} \right] . 
\end{eqnarray}
The first line of \eqref{b-yukawa-a} is simply the result given in appendix D of \cite{heras} multiplied by a Yukawa exponential term. After taking the curl of the first line in \eqref{b-yukawa-a}, taking the limit $\epsilon \to 0$ and converting back to spherical polar coordinates the first line becomes $g \frac{e^{-m r}}{r^2} {\bf {\hat r}} \pm 4 \pi g e^{-mr} \delta (x) \delta (y) \Theta (\mp z) {\bf {\hat z}}$.  The term in the second line of \eqref{b-yukawa-a}, again after taking the limit $\epsilon \to 0$ and converting back to spherical polar coordinates, is $m g \frac{e^{-m r}}{r} \left[\frac{\pm 1 - \cos \theta }{\sin \theta} \right] {\bf {\hat \theta}}$, with ${\bf {\hat \theta}} = -\frac{1}{\sin \theta} ({\bf {\hat z}} - \cos \theta {\bf {\hat r}})$. Combining this results gives the total B-field as
\begin{eqnarray}
\label{b-yukawa-1}
{\bf B} _\pm &=& g \frac{e^{-m r}}{r^2} \left( {\bf {\hat r}} + m r \left[\frac{\pm 1 - \cos \theta }{\sin \theta} \right] {\bf {\hat \theta}} \right)
\pm 4 \pi g e^{-m|z|} \delta (x) \delta (y) \Theta (\mp z) {\bf {\hat z}}~ \nonumber \\
 &=& g \frac{e^{-m r}}{r^2} \left( {\bf {\hat r}} + m r \left[\frac{\pm 1 - \cos \theta }{\sin \theta} \right] {\bf {\hat \theta}} \right)\pm \frac{2 g e^{-m|z|} \delta (\rho)}{\rho}  \Theta (\mp z) {\bf {\hat z}}~.
\end{eqnarray}
The ${\bf \hat r}$ and ${\bf \hat \theta}$ components in ${\bf B}_\pm$ come simply taking the curl of \eqref{b-yukawa} naively, while the last, ${\bf \hat z}$ term is the contribution of the string and arises from the regularization procedure of appendix D of \cite{heras}. In the second line of \eqref{b-yukawa-1} we have used $\delta (x) \delta(y) = \frac{\delta (\rho)}{2 \pi \rho}$ to convert the string part of the magnetic field into cylindrical coordinates. For this last term we have used $e^{-mr} \delta (x) \delta (y) \to e^{-m|z|} \delta (x) \delta (y)$ through the action of the two $\delta$-functions. It is straightforward to see that \eqref{b-yukawa} and \eqref{b-yukawa-1} are exact solutions to the two magnetic equations \eqref{max-mono-m} and \eqref{max-mono-m1} if the $\pm 4 \pi g e^{-m|z|} \delta (x) \delta (y) \Theta (\mp z) {\bf {\hat z}}$ piece of the magnetic field is neglected.   

Next we want to verify the $\nabla \cdot {\bf B}_\pm =0$. Due to the singularities in ${\bf B}_\pm$ from \eqref{b-yukawa-1} it is easier to do this via the integral form of the divergence theorem ({\it i.e.} $\int \nabla \cdot {\bf B}_\pm d^3 x = \int {\bf B}_\pm \cdot d {\bf a} = 0$) rather than directly calculating $\nabla \cdot {\bf B}_\pm$. Taking our surface to be a sphere of radius $R$ the surface integral of the first term in \eqref{b-yukawa-1} is
\begin{eqnarray}
\label{div-b-1}
\int g \frac{e^{-m r}}{r^2} {\bf {\hat r}} \cdot d {\bf a} = g \int \frac{e^{-m R}}{R^2} {\bf {\hat r}} \cdot {\bf {\hat r}} R^2 d \Omega = 4 \pi g e^{-m R}~,
\end{eqnarray}   
where the integration of the solid angle is $\int d \Omega = 4 \pi$. The second term in \eqref{b-yukawa-1} gives zero since ${\bf {\hat \theta}} \cdot d{\bf a} \propto {\bf {\hat \theta}} \cdot {\bf \hat r} = 0$.
Next we look at the string contribution {\it i.e.} the last term in \eqref{b-yukawa-1}. Let us look at the case when the string is along the $-z$-axis for which the string part of the magnetic field in \eqref{b-yukawa-1} is $\frac{+2 g e^{-m|z|} \delta (\rho)}{\rho} \Theta (- z) {\bf {\hat z}}$. This magnetic field will puncture the sphere along the negative $z$-axis at $z=-R$ and due to the $\delta (\rho)$ part we only need to worry about a small area patch, $\pi (\Delta \rho)^2$, located at $z=-R$ and parallel to the $xy$ plane so that $d{\bf a} \approx - {\bf \hat z} \rho d \varphi d \rho$. Since the normal to the sphere is in the ${\bf \hat r}$ direction along the negative $z$ axis one has ${\bf \hat r} = - {\bf \hat z}$ at this location. With this background the surface integral of the string part of the magnetic field along the negative $z$-axis, ${\bf B}_+$, is 

\begin{equation}
\label{div-b-2}
\int \frac{2 g e^{-mR} \delta (\rho)}{\rho} \Theta (- z) {\bf {\hat z}}\cdot d {\bf a} 
= 2 g e^{-m R} \int _0 ^{\Delta \rho} \int _0 ^{2 \pi} \frac{\delta (\rho)}{\rho} {\bf \hat z} \cdot (-{\bf \hat z} \rho d \rho d \varphi) = - 4 \pi g e^{-mR}~.
\end{equation}

Combining \eqref{div-b-1} and \eqref{div-b-2} shows that $\int {\bf B} _+ \cdot d{\bf a} =0$ which by the divergence theorem gives $\int \nabla \cdot {\bf B} _+ d ^3 x =0$ and thus implies $\nabla \cdot {\bf B}_+ =0$. When the string term is neglected one has $\nabla \cdot ({\bf B}_+ - {\bf B}_{string} ) = 4 \pi g \delta ^3 ({\bf r})$ {\it i.e.} a magnetic monopole. The same type of calculation carries through similarly for the case when the string is along the positive $z$-axis {\it i.e.} for $- 4 \pi g e^{-m|z|} \delta (x) \delta (y) \Theta (+ z) {\bf {\hat z}} = - \frac{2 g e^{-m|z|} \delta (\rho)}{\rho}\Theta (+ z) {\bf {\hat z}}$. 

Next we move to the curl of the magnetic field which is 
\begin{eqnarray}
\label{b-curl}
\nabla \times {\bf B}_\pm &=& - m^2 g \frac{ e^{-m r}}{r} \left( \frac{\pm 1 - \cos \theta }{\sin \theta} \right) {\bf {\hat \varphi}} \pm 4 \pi g \Theta (\mp z) e^{-m |z|} \left[ \delta (x) \delta ' (y) {\bf \hat x} - \delta ' (x) \delta (y) {\bf \hat y} \right] \nonumber \\
&=& - m^2 g \frac{ e^{-m r}}{r} \left( \frac{\pm 1 - \cos \theta }{\sin \theta} \right) {\bf {\hat \varphi}} \pm 2 g \Theta (\mp z) e^{-m |z|} \left[ \frac{\delta (\rho)}{\rho^2} -\frac{\delta ' (\rho)}{\rho} \right]{\bf \hat \varphi}~. 
\end{eqnarray}
In the second line we have written the string contribution in cylindrical coordinates.
Combining the first ${\bf \hat \varphi}$ term in \eqref{b-curl} with ${\bf A}_\pm$ from \eqref{b-yukawa} multiplied by $m^2$ we see that these terms cancel and thus \eqref{max-mono-m} is satisfied, except for the string term. As in the massless case, we are left with the current density of the string 
\begin{eqnarray}
\label{j-string}
{\bf J}_{string} &=& \pm g \Theta (\mp z) e^{-m |z|}\left[ \delta (x) \delta ' (y) {\bf \hat x} - \delta ' (x) \delta (y) {\bf \hat y} \right] \nonumber \\
&=& \pm  \frac{g}{2 \pi} \Theta (\mp z) e^{-m |z|} \left[ \frac{\delta (\rho)}{\rho^2} -\frac{\delta ' (\rho)}{\rho} \right]{\bf \hat \varphi} ~.
\end{eqnarray}
This current density is generally not discussed directly, with the exception of works like \cite{shnir} \cite{adorno} \cite{heras}.

There are two points worth commenting on in relation to the above closed form Yukawa-Dirac string solution:
 \begin{enumerate}
 
     \item  The potential for a magnetic charge with a massive photon is obtained in exactly the same way as in the case of electric charge where one takes the Coulomb potential, $\phi = \frac{q}{r}$, and multiplies it by $e^{-mr}$ to obtain the Yukawa potential, $\phi = q \frac{e^{-mr}}{r}$. 
     
     \item The string singularity along the negative/positive axis of ${\bf A}^{(0)} _\pm$, for the massless photon case, is a gauge artifact, since ${\bf A}^{(0)} _+$ and ${\bf A}^{(0)} _-$ are related by a gauge transformation. For the massive case ${\bf A} _+$ and ${\bf A} _-$ are no longer related to one another by a gauge transformation. %The associated magnetic field, has now become a real, physical singularity of the magnetic field, ${\bf B} = \nabla \times {\bf A}^{(0)} _\pm = g\frac{{\bf r}}{r^3}$,  has only the Coulomb singularity at $r=0$. In the massive photon case not only ${\bf A}_\pm$ but the associated magnetic field has a string singularity due to the $\frac{\pm 1 - \cos \theta}{\sin \theta}$ term -- see \eqref{b-yukawa} and \eqref{b-yukawa-1}.  
 \end{enumerate}
 Point number 2 has major consequences for the Dirac quantization condition \cite{dirac,dirac1} between a magnetic charge $g$ and electric charge $q$ which is 
 \begin{equation}
     \label{dirac-cond}
     q g = n \frac{\hbar}{2} ~.
 \end{equation}
One way of arriving at the condition in \eqref{dirac-cond} is by requiring that the Aharonov-Bohm (AB) phase \cite{ab} associated with a charge $q$ moving in the background vector potential ${\bf A}_\pm ^{(0)}$ be undetectable {\it i.e.} that the AB phase be some integer multiple of $2 \pi$ \cite{ryder,heras}. However, now the the string singularity is not only a feature of the 3-vector potential, but is also a feature of the magnetic field. Thus one can not require that the effect of AB phase vanish, since now one has a physical string/solenoid.

There is another way of obtaining the quantization condition of \eqref{dirac-cond} -- the fiber bundle approach of Wu and Yang \cite{wu-yang} where one defines the vector potential in a way that it is non-singular in the region that it covers {\it e.g.} the vector potential is defined as ${\bf A}^{(0)} _+$ over the northern hemisphere ($0<\theta \le \pi/2$) and ${\bf A}^{(0)} _-$ over the southern hemisphere ($\pi/2 < \theta \le \pi$). The two vector potentials are related by the gauge transformation $ {\bf A}^{(0)}_- = {\bf A}^{(0)}_+ - \nabla (\lambda)$ with $\lambda = 2 g \varphi$. This implies that the wavefunctions over these two hemispheres are related by $\Psi _+ = e^{iq \lambda} \Psi_- \to \Psi _+ = e^{2iqg \varphi} \Psi_-$. Since the wavefunction must be single value as $\varphi \to 2 \pi$ this means the phase should be some integer multiple of $2 \pi$ {\it i.e.} $4 \pi q g = 2 \pi n$ which immediately yields the quantization condition from \eqref{dirac-cond}. This Wu-Yang fiber bundle approach also fails in the case of a massive photon since, unlike ${\bf A}^{(0)} _+$ and ${\bf A}^{(0)} _-$, the Yukawa-Dirac vector potentials,  ${\bf A} _+$ and ${\bf A} _-$ from \eqref{b-yukawa}, are not related by a gauge transformation. 

\section{Modified Dirac quantization condition}

One of the most interesting outcomes of Dirac's formulation of magnetic charge in terms of the string potential, ${\bf A}^{(0)} _\pm$, is the existence of a quantization condition between electric and magnetic charges given in \eqref{dirac-cond} for the massless photon case. As shown in the preceding section, when the photon is massive the Dirac quantization condition \eqref{dirac-cond}, can not be obtained using the method of hiding the AB phase associated with the Dirac string or the Wu-Yang fiber bundle method. There is a third approach to the Dirac quantization condition \cite{saha,saha1,wilson} which involves requiring the quantization of the field angular momentum between a magnetic charge $+g$ and an electric charge $+q$. This method {\it might} work. In the massless photon case the field angular momentum of an electric charge-magnetic charge system is \cite{jackson}
\begin{equation}
\label{ang3d}
{\bf L}_{EM} ^{(0)} = \frac{1}{4 \pi} \int {\bf r} \times ({\bf E} \times {\bf B}) d^3 x =  - q g {\bf {\hat R}}~,
\end{equation} 
where ${\bf {\hat R}}$ is the unit vector from the magnetic charge to the electric charge. We have reserved the relative locations of the magnetic and electric charges from that used in \cite{jackson} since we place the magnetic charge at the origin (this is because in the massive photon case the magnetic field is more complex and thus the calculations are easier with the magnetic charge at the origin). Requiring that the magnitude of the angular momentum, ${\bf L}_{EM} ^{(0)}$, be some integer multiple of $\frac{\hbar}{2}$ immediately gives \eqref{dirac-cond}. 

The field angular momentum method of obtaining \eqref{dirac-cond} has been used to study the Dirac quantization condition in the case of a massive photon \cite{joshi}. In addition to reference \cite{joshi} there are a host of other works \cite{joshi2,cafaro,guimaraes,goldhaber,goldhaber2}  which look the quantization condition when the photon is massive. These works come to different and contradictory conclusions, and in fact as noted in \cite{goldhaber} sometimes the same author came to differing conclusions about the possibility of the viability of the Dirac quantization condition in the presence of a photon mass (compare the conclusions of references \cite{goldhaber} and \cite{goldhaber2} on this point). One of the reasons for this is that the string potential and associated magnetic field have, up until now, been incorrectly given. In the section above we have corrected this problem and given the correct string potential \eqref{b-yukawa} and magnetic field \eqref{b-yukawa-1}. We can use the corrected potential and field to investigate the question of what happens to the Dirac quantization condition when the photon is massive. 

With a massive photon the electric charge-magnetic charge  system still carries a field angular momentum, but with some changes. First, the momentum density (the time-space component of the energy-momentum tensor, $T^{0i}$) is altered. For the massless photon case $T^{0i} = ({\bf E} \times {\bf B})^i$, while for the massive photon case $T^{0i} = ({\bf E} \times {\bf B})^i + m^2 \phi {\bf A} ^i$ (see chapter 12 of Jackson's $3^{rd}$ edition ~\cite{jackson3}, in particular problem 12.16). Thus the angular momentum density becomes  $\frac{1}{4 \pi} \left(  {\bf r} \times ({\bf E} \times {\bf B}_\pm) + m^2 \phi {\bf r} \times {\bf A}_\pm  \right)$. To take account of the electric field and potential of the electric charge $q$ we again assume that the charges are fixed so that $\rho _e= q \delta ^3 ({\bf r} - {\bf R})$ and  $\rho _m= g \delta ^3 ({\bf r})$, where ${\bf R}$ is the location of $q$ and we have assumed the magnetic charge is at the origin. Since the charges are at rest the current densities are zero ${\bf J}_e = {\bf J}_m = 0$. This implies the fields are time-independent, $\partial _t {\bf E} = \partial _t {\bf B} =0$. Thus equations \eqref{max-mono-m} and \eqref{max-mono-m1} for ${\bf B}$ and ${\bf A}$ are supplemented by $\nabla \cdot {\bf E} + m ^2 \phi = 4 \pi \rho _{e}  = 4 \pi q \delta ^3 ({\bf r} - {\bf R})$ and $\nabla \times {\bf E}  = 0$ for ${\bf E}$ and $\phi$. The solution to these two additional equations is 
\begin{equation}
\label{e-yukawa}
\phi = q \frac{e^{-m r'}}{r'} ~~~~~~;~~~~~ {\bf E} = -\nabla \phi =q \left( 1 + m r' \right) \frac{e^{-m r'}}{{r'}^2} {\bf {\hat r'}} ~,
\end{equation}
where for convenience we have defined ${\bf r}' = {\bf r} - {\bf R}$. Note that $\phi$ is of the usual Yukawa form.   

We have not been able to work out the general case of the field angular momentum with a massive photon, since the presence of the physical string greatly complicates the calculation. 
There are two special cases where the field angular momentum can be worked out exactly: (i) when the electric charge and magnetic charge are both at the origin; (ii) when the displacement vector between the electric charge and magnetic charge is lined up with the physical Dirac string {\it e.g.} if both the displacement vector between the charges and the physical string lie along the $z$-axis. In this work we make this assume simply for ease of calculation ({\it i.e.} we have not been able to work out the general case analytically). However, there may be some argument to constrain the electric charge to be inside the current density ${\bf J}_{string}$. To investigate this possibility one would need to not put the various electric and magnetic source terms in by hand, but would need to make them dynamical by associating them with some field coupled to the electromagnetic gauge field. We are investigating this possibility, but finding such field interacting solutions is difficult.      

We begin with the first special case when both the electric charge and magnetic charge are located at $r=0$ and we take the string to be along either the $+z$ or $-z$ axis. Using the fields and potentials from \eqref{e-yukawa} \eqref{b-yukawa-1} and \eqref{b-yukawa} the field angular momentum density is
\begin{equation}
\label{ang-mom-den}
\frac{1}{4 \pi } \left( {\bf r} \times ({\bf E} \times {\bf B}_\pm) + m^2 \phi {\bf r} \times {\bf A}_\pm \right) = - \frac{q g m}{4 \pi} (1 + 2mr) \frac{e^{-2 m r}}{r^2} \left[\frac{\pm 1 - \cos \theta }{\sin \theta} \right] {\bf {\hat \theta}} ~.
\end{equation}
Recalling that ${\bf {\hat \theta}}= \cos \theta \cos \varphi {\bf {\hat x}}+ \cos \theta \sin \varphi {\bf {\hat y}} -  \sin \theta {\bf {\hat z}}$ the $d \varphi$ integration of the volume integral of ${\bf r} \times ({\bf E} \times {\bf B}) + m^2 \phi_e {\bf r} \times {\bf A}_\pm$ will give zero for the ${\bf {\hat x}}, {\bf {\hat y}}$ components and $2 \pi$ for the ${\bf {\hat z}}$. Performing the $dr$ and $d \theta$ integrals (we change $d\theta \to dx$ via the substitution $x=\cos \theta$) gives
\begin{eqnarray}
\label{ang-mom}
\frac{1}{4 \pi}  \int \left( {\bf r} \times ({\bf E} \times {\bf B}_\pm) + m^2 \phi {\bf r} \times {\bf A}_\pm \right)  d^3 x 
&=& \frac{q g m}{2} \int _0 ^\infty (1 + 2 m r) e^{-2 m r} dr \int _{-1} ^{1} ( \pm 1 - x ) d x {\bf {\hat z}} \nonumber \\
&=& \pm qg {\bf {\hat z}}  ~.
\end{eqnarray}
On the surface this result is exactly the same as in massless photon case. However, in the usual massless photon case if the electric and magnetic charge are both placed at $r=0$ one gets zero field angular momentum when the chagres are placed at the same location. First, for the radial parts of the electric and magnetic fields one has ${\bf E} ^{(0)}_r \times {\bf B} ^{(0)}_{\pm ~ r} \propto {\bf {\hat r}} \times {\bf {\hat r}} = 0$ which gives ${\bf r} \times ({\bf E}^{(0)}_r \times {\bf B}^{(0)}_{\pm~r}) =0$. Second, there is also a contribution from the from the radial electric field and the string part of the magnetic field as given by the $\pm 4 \pi g \delta(x) \delta (y) \Theta (\mp z) {\bf {\hat z}}$ in \eqref{regular-B2}. In this case the radial part of the electric field and this string piece give  ${\bf E} ^{(0)}_r \times {\bf B} ^{(0)}_{\pm ~ string} \propto \delta(x) \delta (y) {\bf {\hat r}} \times {\bf {\hat z}} = - \delta(x) \delta (y) \sin \theta {\bf {\hat \varphi}}$, where we have used the fact that ${\bf {\hat z}} = {\bf {\hat r}} \cos \theta -{\bf {\hat \theta}} \sin \theta$, ${\bf {\hat r}} \times {\bf {\hat r}} = 0$ and ${\bf {\hat r}} \times {\bf {\hat \theta}} = {\bf {\hat \varphi}}$. Finally, using all this one finds that  ${\bf r} \times ({\bf E}^{(0)}_r \times {\bf B}^{(0)}_{\pm~string}) \propto  - \delta(x) \delta (y) \sin \theta  ~ {\bf {\hat r}} \times {\bf {\hat \varphi}} =\delta(x) \delta (y) \sin \theta  ~ {\bf {\hat \theta}}$. Since $\sin \theta = \frac{\sqrt{x^2+y^2}}{r}$ then the two delta functions, $\delta(x) \delta (y)$ will make $\sin \theta \to 0$ and thus the entire integrand and integral go to zero. Thus the string piece of the magnetic field will not contribute to the field angular momentum if the electric charge sits on the same axis at the string.  

Returning now to the massive photon case for the special case when both the electric charge and magnetic charge are at the origin, we see that the field angular momentum in \eqref{ang-mom} comes from the vector potential piece $m^2 \phi {\bf r} \times {\bf A}_\pm$, and from the ${\bf \hat \theta}$ part of the magnetic field in ${\bf r} \times ({\bf E} \times {\bf B}_\pm)$. This field angular momentum in \eqref{ang-mom} points along $+{\bf {\hat z}}$ if the singularity is along the $-z$-axis, and points along $-{\bf {\hat z}}$ if the singularity is along the $+z$-axis. In this special case one recovers the standard quantization condition of \eqref{dirac-cond} in agreement with references \cite{cafaro,guimaraes,goldhaber} and in disagreement with reference \cite{joshi,joshi2,goldhaber2}.

We now move to the second special case -- having the string direction align with the displacement between the charges. We will take the magnetic charge to be located at $r=0$ and we will first consider the case when the physical string lies along the $-z$ axis {\it i.e.} we will use the ${\bf B}_+$ and ${\bf A}_+$ fields from \eqref{b-yukawa-1} and \eqref{b-yukawa} respectively. The magnetic charge is placed at the origin since it is the more complicated field. Next we place the electric charge on the $\pm z$-axis a distance $R$ from the origin so that its position is given by $\pm R {\bf \hat z}$. For $- R {\bf \hat z}$ the string will run through the charge $q$, while for $+ R {\bf \hat z}$ the charge $q$ will be located outside the string. The electric potential and electric field for $q$ for these locations is given by \eqref{e-yukawa} with ${\bf r'} = {\bf r} \mp R {\bf \hat z}$ depending if the electric charge is at $+ R {\bf \hat z}$ or $- R {\bf \hat z}$. The magnitude of the position vector from the electric charge is $r'=\sqrt{r^2 + R^2 \mp 2rR \cos \theta}$. We present the details of the calculation of the field angular momentum in the appendix. The summary of these calculations from the appendix is 

\begin{enumerate}[i]
    
\item {\it String along $-z$-axis, electric charge at $r=-R{\bf \hat z}$:} ${\bf L}_{EM} = {\bf L}_{EM} ^{point} + {\bf L}_{EM} ^{A+B^\theta} = 2 qg e^{-m R} {\bf \hat z}$ 

\item {\it String along $-z$-axis, electric charge at $r=+R{\bf \hat z}$:}  ${\bf L}_{EM} = {\bf L}_{EM} ^{point} + {\bf L}_{EM} ^{A+B^\theta} = 0$

\item {\it String along $+z$-axis, electric charge at $r=-R{\bf \hat z}$:} ${\bf L}_{EM} = {\bf L}_{EM} ^{point} + {\bf L}_{EM} ^{A+B^\theta} = 0$ 

\item {\it String along $+z$-axis, electric charge at $r=+R{\bf \hat z}$:} ${\bf L}_{EM} = {\bf L}_{EM} ^{point} + {\bf L}_{EM} ^{A+B^\theta} = -2 qg e^{-mR} {\bf \hat z}$ 

\end{enumerate}

The usual result for the field angular momentum in the case of a massless photon is given in \eqref{ang3d} as ${\bf L}^{(0)} _{EM} = qg {\bf \hat R}$. The results above -- minus the factor $e^{-mR}$ -- seem to indicate either twice this value (for cases (i) and (iv)) or $0$ (for cases (ii) and (iii)). Cases (i) and (iv) may be understood as the addition of the point and string contributions to give twice the result of the massless case, $2 qg$, with a multiplicative $e^{-mR}$ factor to account for the photon mass. Cases (ii) and (iii) may be understood as a cancellation of the point and string contributions. Note that in the above summary of the results for ${\bf L}_{EM}$ we have included the contribution of the magnetic field coming from the point/Yukawa part, the vector potential part and the ${\bf \hat \theta}$ component, but we have apparently neglected the contribution coming from the string part of the magnetic field ({\it e.g.} the ${\bf \hat z}$ term in \eqref{b-yukawa-1}). As explained in the appendix the contribution from the string part of the magnetic field vanishes when the electric charge sits on the same axis as the string.  

If we now apply the requirement that ${\bf L}_{EM}$ from above must be some integer multiple of $\hbar / 2$ we would get conditions from cases (i) and (iv) like
\begin{equation}
    \label{dirac-mass}
    2 qg e^{-mR} = n \frac{\hbar}{2} ~.
\end{equation}
This quantization condition involves not only the charges, $q, g$, but also involves mass and position, $m$ and $R$. To check more carefully whether the heuristic arguments leading to \eqref{dirac-mass} are correct, one should calculate the commutators of the total angular momentum operators \footnote{One should check that $[L_i, L_j] = i \hbar \epsilon _{ijk} L_k$ where $L_i = L_i ^{particle} + L_i ^{field}$} to see if these work out and, if so, under what conditions. The calculation of the commutators for the total angular momentum (particle angular momentum plus field angular momentum) of the electric charge-magnetic charge system for the massless photon case was carried in \cite{fierz,lipkin,yang}. If the particle carrying either the electric charge or the magnetic charge has an intrinsic spin then this should be included in the total angular momentum as ${\bf L}_{tot} = {\bf L}_{spin} + {\bf L}_{orbital} + {\bf L}_{field}$. It is this ${\bf L}_{tot}$ which should satisfy the angular momentum commutation relationship. 

A system with a field angular momentum that is closer to the present case of electric charge-magnetic charge plus photon mass is the field angular momentum of an electric charge, $q$, plus a magnetic dipole, ${\bf m}$, with a massless photon. Such a system of electric charge-magnetic dipole was investigated in \cite{singleton-plb} as a way of addressing the nucleon-spin puzzle. In \cite{singleton-98} it was shown that the total angular momentum of the electric charge-magnetic dipole system exactly satisfied the standard commutator for angular momentum. We are currently studying the general case of the field angular momentum of the electric charge-magnetic charge system with a massive photon and the viability (or not) of the Dirac quantization condition. One similarity between the electric-charge magnetic dipole system with the electric-charge magnetic-charge plus massive photon system is that both have a special direction -- in the first case it is the direction of the magnetic dipole, ${\bf m}$, and in the second case it is the direction of the magnetic string.

\section{Summary and Conclusion}

In this work we have given the Dirac string potential and magnetic field, equations \eqref{b-yukawa} and \eqref{b-yukawa-1}, for a electromagnetism with a magnetic charge plus massive photon. This corrects the previous results for this system given in reference \cite{joshi}. The vector potential in \eqref{b-yukawa}, is similar to what occurs for the Yukawa scalar potential for electric charge given in \eqref{e-yukawa} -- one takes the $m=0$ potential (either Coulomb scalar potential or Dirac string potential) and multiplies it by the Yukawa factor $e^{-mr}$. This was shown by explicit calculations using the Maxwell equations with magnetic charge and photon mass given by  \eqref{e-maxwell3-source} \eqref{b-maxwell3-source-1} \eqref{b-maxwell3-source} \eqref{e-maxwell3-source-1}.  

Combining photon mass and magnetic charge turned the Dirac string from a gauge artifact into a real, physical string, and spoiled the spherical symmetry of the magnetic field associated with the magnetic charge due to the addition of the ${\bf \hat \theta}$ dependent string part in \eqref{b-yukawa-1}. Thus adding a photon mass not only spoils the gauge symmetry but also spoils the radial symmetry of the magnetic field. This connection between gauge and spatial symmetries warrants further investigation.  

In section II we investigated the fate of the Dirac quantization condition in the presence of a photon mass. Since the string singularity went from a gauge artifact, in the massless photon case, to a real, physical feature in the massive photon case, the approach to the Dirac quantization condition which required the vanishing of the Aharonov-Bohm phase of the string \cite{ryder} no longer worked. Also the fiber bundle approach \cite{wu-yang} to the Dirac quantization condition was no longer viable. However the method of obtaining the Dirac quantization condition through the quantization of the field angular momentum of the electric charge - magnetic charge system \cite{saha,saha1,wilson} still gave a potential path toward finding some form of the Dirac quantization condition when $m \ne 0$. We were not able to obtain a closed form expression for ${\bf L}_{EM}$ in the general case, but we were able to investigate two special cases: (i) when the electric charge and magnetic charge where at the same location and (ii) when the displacement vector between the charges was aligned along the direction of the string. In the first case we did recover the original Dirac quantization condition \eqref{dirac-cond}, and in the second case we obtained a modified variant of the Dirac condition given in equation \eqref{dirac-mass} which involved both $m$ and $R$. We are currently working on using the correct vector potential and magnetic fields from \eqref{b-yukawa} and \eqref{b-yukawa-1} to perform an analysis similar to references \cite{fierz,lipkin,yang} by looking at how the commutation relationship for the full angular momenta, ( {\it i.e.} $[L_i, L_j] = i \hbar \epsilon _{ijk} L_k$) plays out for a massive photon; to see if this is possible or not, and if so what kind of condition needs to be imposed on the electric charges, magnetic charges, and potentially the photon mass and displacement between the charges.  

In this work we have also included the magnetic field and current density source connected with the Dirac string -- see equations  \eqref{b-yukawa-1} and \eqref{b-curl}. These fields and sources for the Dirac string are not generally emphasized, but recent works \cite{shnir} \cite{adorno} \cite{heras} have pointed out the subtle, but crucial importance of these quantities. In particular one may ask about the nature of the source of the effective string current in \eqref{b-curl}, ${\bf J}_{string} = \pm g \Theta (\mp z) \left[ \delta (x) \delta ' (y) {\bf \hat x} - \delta ' (x) \delta (y) {\bf \hat y} \right]$. Note that this current density has a rotational character and thus might be expected to contribute to the angular momentum of the system. Within the context of the present work we can not address this question since we have put the sources in by hand {\it e.g.} we had taken the magnetic point source to have the form $\rho_m = g \delta ^3 ({\bf r})$ without saying what fields give rise to this source. To properly and consistently take into account the sources, we should include an additional field interacting with the gauge field in the manner of the t' Hooft - Polyakov monopole \cite{thooft} \cite{polyakov} or the Prasad-Sommerfield solution \cite{prasad}. We are currently looking into this possibility of modeling the inserted-by-hand magnetic sources with field sources.

\appendix
%\appendixpage
%\addappheadtotoc

\section{Field angular momentum for displaced electric charge}
In this appendix we present the details of the calculation of the field angular momentum in the special case where the magnetic charge and electric charge are at different points, but with the displacement vector between the charges aligned with physical string. We start with the magnetic field and vector potential with the string along the $-z$ axis given by \eqref{b-yukawa-1} and \eqref{b-yukawa} as
\begin{equation}
\label{b-yukawa-app}
{\bf B} _+ = g \frac{e^{-m r}}{r^2} \left( {\bf {\hat r}} + m r \left[\frac{ 1 - \cos \theta }{\sin \theta} \right] {\bf {\hat \theta}} \right) ~~~~;~~~~ {\bf A}_+ = g \frac{e^{-m r}}{r} \left( \frac{ 1 - \cos \theta }{\sin \theta} \right) {\bf {\hat \varphi}} ~.
\end{equation}
We take the electric field and potential for a charge located on the $z$-axis at $-R$ (inside the string) given by 
\begin{equation}
    \label{e-yukawa-app}
    {\bf E} = q (1 + m r') \frac{e^{-mr'}}{(r')^3} {\bf  r'} ~~~~;~~~~
    \phi = q \frac{e^{-m r'}}{r'} ~,
\end{equation}
where ${\bf r'} = {\bf r} + R {\bf \hat z}$ and $r' =\sqrt{r^2 +R^2 + 2 r R \cos \theta}$. The case when the charge sits at $+R {\bf \hat z}$ (outside the string) has a parallel calculation but with the changes ${\bf r'} = {\bf r} - R {\bf \hat z}$ and $r' =\sqrt{r^2 +R^2 - 2 r R \cos \theta}$. We find that for the former case the field angular momentum of the string-part adds to that of the point-part, while in the latter case the the string-part cancels that of the point-part. In the next two subsections we calculate the contribution to the field angular momentum of the point-part and string-part of the fields in \eqref{b-yukawa-app} and \eqref{e-yukawa-app}.

\subsection{Point-part}

Here we calculate the point-part contribution to the field angular momentum density from the radial part of the magnetic field ${\bf B}^{point} _+ = g \frac{e^{-m r}}{r^2} {\bf {\hat r}}$. The double cross product from  ${\bf r} \times ({\bf E} \times {\bf B}^{point} _+)$ works out as ${\bf r} \times ({\bf r'} \times {\bf \hat r}) = R \sin \theta {\bf r} \times {\bf \hat \varphi} = - r R \sin \theta {\bf \hat \theta}$. The field angular momentum for this point-part is now
\begin{equation}
    \label{point-1}
    {\bf L}_{EM} ^{point} = - \frac{R q g}{4 \pi} \int \frac{\sin \theta (1 + m r')e^{-mr'}e^{-mr}}{r {r'}^3} {\bf \hat \theta} d^3 x ~.
\end{equation}
We can perform the integration over $d \varphi$ since the only $\varphi$ dependence is in ${\bf \hat \theta}$. This yields $\int _0 ^{2 \pi} {\bf \hat \theta} d \varphi = -2 \pi \sin \theta {\bf \hat z}$. Collecting terms and making the standard change of variables $x=\cos \theta$ we obtain
\begin{equation}
    \label{point-2}
    {\bf L}_{EM} ^{point} =  \frac{R q g}{2} {\bf \hat z} \int _0 ^\infty \int _{-1} ^1 \frac{(1-x^2) (1 + m r')e^{-mr'}e^{-mr}}{r {r'}^3} ~ dx ~ r^2 dr ~.
\end{equation}
Using {\it Mathematica} the $dx$ integration yields (taking into account $r'=\sqrt{r^2+R^2+2rRx}$)
\begin{eqnarray}
    \label{point-3}
    {\bf L}_{EM} ^{point} =  \frac{q g}{R^2 m^3} {\bf \hat z} \int _0 ^\infty \frac{e^{-mr}}{r^2} \bigg( && e^{-m |r-R|} (-1 +m^2rR -m|r-R|) \nonumber \\
    &&+ e^{-m(r+R)} (1 + m^2rR + m (r+R)) \bigg)  ~ dr ~.
\end{eqnarray}
Due to the presence of the $|r-R|$ term, the $dr$ integration has to be broken up into the range $0 \le r \le R$ (when $|r-R| = R-r$) and $R \le r \le \infty$ (when $|r-R| = r-R$). Using {\it Mathematica} the $dr$ integration yields
\begin{equation}
    \label{point-4}
    {\bf L}_{EM} ^{point} =  \frac{q g e^{-m R}}{m^2 R^2} \left( -2 m R + (1 + m R)(\gamma + \ln (2 m R)) + e^{2 m R} (-1+mR) Ei (-2 m R) \right) {\bf \hat z}~.
\end{equation}
In \eqref{point-4} $\gamma \approx 0.577216$ is the Euler–-Mascheroni constant and $Ei (x) = - \int _{-x} ^\infty \frac{e^{-t}}{t} dt$ is the exponential integral function.

\subsection{${\bf A}$ and $B^{\theta}$ parts}

Here we calculate the contribution to the field angular momentum density from those parts of the vector potential and magnetic field that are proportional to $\left( \frac{ 1 - \cos \theta }{\sin \theta} \right)$. From \eqref{b-yukawa-app} one can see that there are two contributions: ${\bf B}^{\theta} _+ = g m \frac{e^{-m r}}{r} \left[\frac{ 1 - \cos \theta }{\sin \theta} \right] {\bf {\hat \theta}}$ and ${\bf A}_+ = g \frac{e^{-m r}}{r} \left( \frac{ 1 - \cos \theta }{\sin \theta} \right) {\bf {\hat \varphi}}$. We will calculate the latter term first.  

The contribution to the field angular momentum density from ${\bf A}_+$ is $\frac{m^2}{4 \pi} \phi ( {\bf r} \times {\bf A}_+)$. The cross product yields ${\bf \hat r} \times {\bf \hat \varphi} = - {\bf \hat \theta}$. As previously the $d \varphi$ integration gives $\int _0 ^{2 \pi} {\bf \hat \theta} d \varphi = - 2 \pi \sin \theta {\bf \hat z}$. Using $\phi$ and ${\bf A}_+$ from \eqref{e-yukawa-app} and \eqref{b-yukawa-app} respectively we find
\begin{equation}
    \label{stringA-1}
    {\bf L}_{EM} ^{A} = \frac{m^2 q g}{2} {\bf \hat z}\int _0 ^\infty r^2 dr \int _{-1} ^1 dx \frac{e^{-mr} e^{-mr'}(1-x)}{r'} ~.
\end{equation}
We have made the change of variables $x=\cos \theta$. Using {\it Mathematica} to carry out the $dx$ integration yields
\begin{eqnarray}
    \label{stringA-2}
    {\bf L}_{EM} ^{A} = \frac{q g}{2m R^2} {\bf \hat z}\int _0 ^\infty  dr e^{-mr} \bigg(&& -e^{-m|r-R|}(1+m|r-R| -2 m^2 r R) \nonumber \\
    &&+ e^{-m(r+R)} (1 + m (r+R)) \bigg) ~.
\end{eqnarray}
Due to the $|r-R|$ term one again needs to split the $dr$ integration up into two ranges -- $0\le r \le R$ and $R \le r \le \infty$. Using {\it Mathematica} to do the $dr$ integration yields
\begin{equation}
    \label{stringA-3}
    {\bf L}_{EM} ^{A} = \frac{1}{4} q g e^{-mR} (1 + 2 m R) {\bf \hat z} ~.
\end{equation}

Now we move on to  ${\bf B} ^{\theta} _+$. We first compute the double cross product ${\bf r} \times ({\bf r'} \times {\bf \hat \theta} ) = {\bf r} \times (r + r \cos \theta ) {\bf \hat \varphi} = - r(r + R \cos \theta) {\bf \hat \theta}$. As before the only $\varphi$ dependence comes from ${\bf \hat \theta}$ so we can perform the $d \varphi$ integration $\int ^{2 \pi} _0 {\bf \hat \theta} d \varphi = -2 \pi \sin \theta {\bf \hat z}$. Collecting terms and making the change of variable $x=\cos \theta$ we find
\begin{equation}
    \label{stringB-1}
    {\bf L}_{EM} ^{B^\theta} = \frac{m q g}{2} {\bf \hat z}\int _0 ^\infty r^2 dr \int _{-1} ^1 dx \frac{e^{-mr} e^{-mr'}(1-x)(1+mr')(r+Rx)}{(r')^3} ~.
\end{equation}
Carrying out the $dx$ integration via {\it Mathematica} yields
\begin{eqnarray}
    \label{stringB-2}
    {\bf L}_{EM} ^{B^\theta} &=& \frac{q g}{2m^2 R^2} {\bf \hat z}\int _0 ^\infty r^2 dr  \frac{e^{-mr}}{r^3} \bigg( -e^{-m|r-R|} \Big[ 2m |r-R|+2+m^2r(r-3R) \pm 2 m^3r^2 R \Big] \nonumber \\
   &+& e^{-m(r+R)} \Big[ 2m (r+R) +2 +m^2r(r+R) \Big] 
\bigg)  ~.
\end{eqnarray}
The $+2m^3 r^2 R$ term in the first line is for the case when $r<R$ and the $-2m^3 r^2 R$ term in the first line is for the case when $r>R$. As previously the $dr$ integration has to be split into $r<R$ and $r>R$ ranges. Using {\it Mathematica } the $dr$ integration of \eqref{stringB-2} gives 
\begin{eqnarray}
    \label{stringB-3}
    {\bf L}_{EM} ^{B^\theta} = \frac{q g e^{-mR}}{4m^2 R^2}  &\Big(& mR(8+7mR-2 m^2 R^2) 
- 4(1 + mR) (\gamma + \ln (2 mR) )  \nonumber \\
&-&4e^{2mR} (-1 +mR) Ei(-2 m R)  \Big){\bf \hat z} ~.
\end{eqnarray}
As before $\gamma$ is the Euler-–Mascheroni constant and $Ei (x)$ is the exponential integral function. 
Adding together the two parts from \eqref{stringB-3} and \eqref{stringA-3} gives the total string contribution from the case when the string is along the $-z$-axis and the electric charge is embedded in the string at $r=-R{\bf \hat z}$ as
\begin{eqnarray}
    \label{stringT}
    &&{\bf L}_{EM} ^{A+B^\theta} = {\bf L}_{EM} ^{A}+{\bf L}_{EM} ^{B^\theta}  \\
    &=&\frac{q g e^{-mR}}{m^2 R^2}  \Big( 2 m R (1+mR)-(1 + mR) (\gamma + \ln (2 mR) ) -e^{2mR} (-1 +mR) Ei(-2 m R)  \Big){\bf \hat z}
  ~. \nonumber
\end{eqnarray}

\subsection{Total field angular momentum}

In this subsection we obtain the total field angular momentum by combining the  results from the previous two subsections for the four different cases: (i) string along the $-z$-axis and electric charge located at $r=-R{\bf \hat z}$; (ii) string along the $-z$-axis and electric charge located at $r=+R{\bf \hat z}$ ; (iii) string along the $+z$-axis and electric charge located at $r=-R{\bf \hat z}$; (iv) string along the $+z$-axis and electric charge located at $r=+R{\bf \hat z}$. Now in the above calculations we have apparently not included the contribution coming from the string part of the magnetic field ({\it i.e.} the $\pm 4 \pi g e^{-m|z|} \delta (x) \delta (y) \Theta (\mp z) {\bf {\hat z}}$ term in \eqref{b-yukawa-1}). However, it is straightforward to check that the field angular momentum from this part of the magnetic field is zero when the electric charge is on the same axis as the string, as is the case when the photon is massless (see the calculation after equation \eqref{ang-mom}).

Case (i) is the case that we explicitly calculated in the above two subsections. Adding together the point part from \eqref{point-4} and string part from \eqref{stringT} gives 
\begin{equation}
    \label{lem--}
    {\bf L}_{EM} = {\bf L}_{EM} ^{point} + {\bf L}_{EM} ^{A+B^\theta} = 2 qg e^{-m R} {\bf \hat z}
\end{equation}

Case (iv) can be obtained from case (i) by simply exchanging the positive and negative $z$-axis. This switch changes the signs of  \eqref{point-4} and \eqref{stringT}. Adding these sign changed terms together the total field angular momentum becomes  ${\bf L}_{EM} = {\bf L}_{EM} ^{point} + {\bf L}_{EM} ^{A+B^\theta} = -2 qg e^{-mR} {\bf \hat z}$ {\it i.e.} the same magnitude but opposite direction from case (i). 

Case (ii) requires that one repeat the calculations of the above two subsections that lead to \eqref{point-4} \eqref{stringA-3} and \eqref{stringB-3} but with ${\bf r}'={\bf r} - R {\bf \hat z}$ and $r'=\sqrt{r^2 + R^2 - 2 r R \cos \theta}$. We do not give the details explicitly but the results are 
\begin{eqnarray}
\label{lem-+}
{\bf L}_{EM} ^{point} &=&  \frac{q g e^{-m R}}{m^2 R^2} \left( 2 m R - (1 + m R)(\gamma +\ln (2mR) ) - e^{2 m R} (-1+mR) Ei (-2 m R) \right) {\bf \hat z}~, \nonumber \\
{\bf L}_{EM} ^{A} &=& \frac{1}{4}qg e^{-mR} {\bf \hat z}~, \\  
{\bf L}_{EM} ^{B^\theta} &=&  \frac{q g e^{-mR}}{4m^2 R^2}  \Big( -mR(8+mR)+4(1 + mR) (\gamma + \ln (2 mR) )  \nonumber \\
&&~~~~~~~~~~~~+ 4e^{2mR} (-1 + mR) Ei(-2 m R)  \Big) {\bf \hat z} ~.\nonumber
\end{eqnarray}
Adding up all the terms in \eqref{lem-+} gives
\begin{equation}
    \label{lem-+2}
    {\bf L}_{EM}  = {\bf L}_{EM} ^{point} + {\bf L}_{EM} ^{A}  + {\bf L}_{EM} ^{B^\theta}  = 0~,
\end{equation}
In this case the different contributions to the field angular momentum cancel. 

Case (iii) can be obtained from case(ii) simply by flipping the $z$-axis around exchanging the positive and negative $z$-axis. This changes the overall sign in front of each term in \eqref{lem-+}. When adding these together one again obtains the result that the field angular momentum is zero, ${\bf L}_{EM} = {\bf L}_{EM} ^{point} + {\bf L}_{EM} ^{A}  + {\bf L}_{EM} ^{B^\theta}  = 0$.

\subsection{Limits and checks of the field angular momentum results}

We now want to check the above result, and come to a physical understanding of the field angular momentum by taking the $m \to 0$ and/or $R \to 0$ limits. In looking at the expression for the point contribution, ${\bf L} ^{point} _{EM}$ (see \eqref{point-4} and \eqref{lem-+}),  and the two string contributions, ${\bf L}_{EM} ^{string-A}$ and ${\bf L}_{EM} ^{string-B}$ (see \eqref{stringA-3}, \eqref{stringB-3} and \eqref{lem-+}) one sees that all of these terms depend on $mR$. Thus the two limits $m \to 0$ and $R \to 0$ are intertwined with one another. Defining the new variable $x=mR$ we can write the point contribution as
\begin{equation}
    \label{limit-point}
    {\bf L}_{EM} ^{point} =  \pm \frac{q g e^{-x}}{x^2} \left( -2 x + (1 + x)(\gamma + \ln (2 x)) + e^{2 x} (-1+x) Ei (-2 x) \right) {\bf \hat z}~,
\end{equation}
where the $+$ sign is for \eqref{point-4} and the $-$ sign is for \eqref{lem-+}. Using {\it Mathematica} and taking $x\to 0$ one finds
\begin{equation}
    \label{limit-point2}
    \lim_{x \to 0} {\bf L}_{EM} ^{point} =  \pm qg  {\bf \hat z}~~~~~{\rm Limit ~~} x, m \to 0 ~~{\rm but} ~~ R \ne 0.
\end{equation}
One must be careful since the above limit only applies for $m\to 0$, but not $R \to 0$. For $R \to 0$ the vector direction of both the electric and magnetic field are solely in the ${\bf \hat r}$ direction. Thus $\lim_{R \to 0} {\bf E} ^{point} \times {\bf B}^{point} \propto {\bf \hat r} \times {\bf \hat r} = 0$ and ${\bf L}_{EM} ^{point} = 0$. This is in agreement with the massless case where ${\bf L}_{EM} ^{(0)} =0$ when the electric and magnetic charges sit on top of one another. In summary the limit in \eqref{limit-point2} does agree with the $m=0$ case {\it if} the electric charge and magnetic charge are displaced from one another. When $R=0$ from the outset, going back to the beginning of the calculation in the appendix and using the fact that ${\bf \hat r} \times {\bf \hat r} = 0$ leads to the consistent result ${\bf L}_{EM}^{point}=0$.

Next we look at the contribution from the vector potential for cases (i) and (ii) when the string is along the $-z$-axis. When the charge lies at $-R{\bf \hat z}$ equation \eqref{stringA-3}  gives ${\bf L}_{EM} ^{A} = \frac{1}{4} q g e^{-x} (1 + 2 x) {\bf \hat z}$; when the charge lies at $+R{\bf \hat z}$ equation \eqref{lem-+} gives ${\bf L}_{EM} ^{A} = \frac{1}{4} q g e^{-x} {\bf \hat z}$. In either case (i) or (ii) one finds that  $\lim_{x \to 0} {\bf L}_{EM} ^{A} = \frac{1}{4} q g {\bf \hat z}$ {\it i.e.} this piece of the field angular momentum is along the $+z$-direction. For cases (iii) and (iv) when the string is placed along the $+z$-axis the limit $x \to 0$ will simply have the opposite sign from cases (i) and (ii). In summary
\begin{equation}
    \label{limit-a}
    \lim_{x \to 0} {\bf L}_{EM} ^{A} =  \pm \frac{1}{4} qg  {\bf \hat z}~~~~~{\rm Limit ~~} x, R \to 0 ~~{\rm but} ~~ m \ne 0 ~,
\end{equation}
with the $+$ sign being for the string along the $-z$-axis and the $-$ sign being for the string along the $+z$-axis. For $x \to 0$ but $m=0$ one finds ${\bf L}_{EM} ^{A}=0$, since this term is missing from the outset -- see \eqref{stringA-1}.

Finally, we tackle the limit of the $\theta$ part of the magnetic field, ${\bf L}_{EM}^{B^\theta}$. For cases(i) and (ii) (given in \eqref{stringB-3} and \eqref{lem-+}) taking the limit $x \to 0$ but $m\ne 0$ one finds that $\lim_{x \to 0}{\bf L}_{EM}^{point-B} = \frac{3}{4} q g {\bf \hat z}$. Running through cases (iii) and (iv) one finds that $\lim_{x \to 0}{\bf L}_{EM}^{point-B} = - \frac{3}{4} q g {\bf \hat z}$. Summarizing this contribution in the limit gives
\begin{equation}
    \label{limit-b}
    \lim_{x \to 0} {\bf L}_{EM} ^{string-B} =  \pm \frac{3}{4} qg  {\bf \hat z}~~~~~{\rm Limit ~~} x, R \to 0 ~~{\rm but} ~~ m \ne 0 ~.
\end{equation}
As with the contribution ${\bf L}_{EM} ^{A}$ the $+$ sign is for the string along the $-z$-axis and the $-$ sign is for the string along the $+z$-axis. For $x \to 0$ but $m=0$ ${\bf L}_{EM} ^{B^\theta}=0$ since this term is missing from the outset -- see \eqref{stringB-1}. 

These limits confirm previous results. If $m=0$ to begin with and then ${\bf L}_{EM} ^{A} = {\bf L}_{EM} ^{B^\theta}=0$ while from \eqref{limit-point2} ${\bf L}_{EM} ^{point} = \pm qg {\bf \hat z}$ which agrees with ${\bf L}_{EM} ^{(0)} = qg {\bf \hat R}$ from \eqref{ang3d}. If $R=0$ but $m \ne 0$ then ${\bf L}_{EM} ^{point} = 0$ while ${\bf L}_{EM} ^{A} = \pm \frac{1}{4} q g {\bf \hat z}$ and $ {\bf L}_{EM} ^{B^\theta}= \pm \frac{3}{4} q g {\bf \hat z}$ from \eqref{limit-a} and \eqref{limit-b}. Adding these together gives ${\bf L}^{A+B^\theta} _{EM} = \pm qg {\bf \hat z}$ which agrees with the result in \eqref{ang-mom}. In this case all the field angular momentum comes from the ${\bf A}$ and ${\bf B}^\theta$ parts.

Finally, these limits also support the four cases when the electric charge is displaced from the magnetic charge, as enumerated below equation \eqref{ang-mom}. For case (i) both the point and ${\bf A}$ plus ${\bf B}^\theta$ contributions point in the $+z$ direction and have magnitudes $qg$ so that the results add for the two parts to give $2 qg {\bf \hat z}$ times an $e^{-mR}$ factor. For case (ii) one changes the location of the electric charge which reverses the direction of ${\bf L}_{EM} ^{point}$ but not the direction of ${\bf L}_{EM} ^{A}$ or $ {\bf L}_{EM} ^{B^\theta}$, so that now the point and ${\bf A}$ plus ${\bf B}^\theta$ contributions cancel giving ${\bf L}_{EM} =0$. Case (iii) is simply case (ii) but with both the string and electric charge flipped along the $z$-axis relative to case (ii). Thus for case (iii) the string and point charges again cancel giving ${\bf  L}_{EM}=0$. Finally, case (iv) is the same as case (i) but with both the string and point contributions flipped relative to the $z$-axis. The string and point charges again add giving ${\bf L}_{EM} = -2 q g {\bf \hat z}$ times an $e^{-mR}$ factor.


\begin{thebibliography}{99}

\bibitem{dirac} P.A.M. Dirac, %``Quantised Singularities in the Electromagnetic Field" , 
Proc. Roy. Soc. A {\bf 133}, 60-72 (1931).

\bibitem{dirac1} P.A.M. Dirac, %``The Theory of Magnetic Poles" , 
Phys. Rev. {\bf 74}, 817-830 (1948).

\bibitem{proca} A. Proca,  J.Phys.Radium {\bf 7}, 347 (1936).  

\bibitem{joshi} A. Yu. Ignatiev and G. C. Joshi, %``Massive electrodynamics and the magnetic monopoles", 
Phys. Rev. D {\bf 53}, 984-992 (1996).

\bibitem{jackson} J.D. Jackson, {\em Classical Electrodynamics}  $2^{nd}$ edition, (John Wiley \& Sons, 1975).

\bibitem{shnir} Y.M. Shnir, {\it Magnetic Monopoles} (Springer, Berlin 2005.

\bibitem{adorno} T.C. Adorno, D.M. Gitman, and A.E. Shabad, Proc. of the Steklov Inst. of Math. {\bf 309}, 1 (2020).

\bibitem{heras} R. Heras, Contemp. Phys. {\bf 59}, 331 (2018).

\bibitem{evans} T. Evans and D. Singleton, Int. J. Mod. Phys. A {\bf 33}, 1850064 (2018).

\bibitem{ab} Y. Aharonov and D. Bohm, %``Significance of electromagnetic potentials in the quantum theory",
Phys. Rev. {\bf 115}, 484 (1959).

\bibitem{ryder} L.H. Ryder, {\em Quantum Theory of Fields} 2$^{nd}$ edition (Cambridge University Press 1996).

\bibitem{wu-yang} T.T. Wu and C,N. Yang, Phys. Rev. D {\bf 12}, 3845 (1975).

\bibitem{saha} M.N. Saha, %``On the Origin of Mass in Neutrons and Proton's", 
Ind. J. Phys. {\bf 10}, 145-151 (1936).

\bibitem{saha1} M.N. Saha, %``Note on Dirac's Theory of Magnetic Poles", 
Phys. Rev. {\bf 75}, 1968 (1949).

\bibitem{wilson} H.A. Wilson, %``Note on Dirac's Theory of Magnetic Poles", 
Phys. Rev. {\bf 75}, 309 (1949).

\bibitem{joshi2} A.Yu. Ignatiev and G. C. Joshi, Mod.Phys.Lett. A {\bf 11}, 2735 (1996)

\bibitem{cafaro} C. Cafaro, S. Capozziello, Ch. Corda, and S.A. Ali, Adv.High Energy Phys., 69835 (2007).

\bibitem{guimaraes} M.S. Guimaraes, R. Rougemont, C. Wotzasek, and C.A.D. Zarro, Phys.Lett.B {\bf 723}, 422 (2013).

\bibitem{goldhaber} A. S. Goldhaber and R. Heras, ``Dirac Quantization Condition Holds with Nonzero Photon Mass", e-Print: 1710.03321 

\bibitem{goldhaber2} A. S. Goldhaber and M. M. Nieto, Rev.Mod.Phys. {\bf 43}, 277 (1971).

\bibitem{jackson3} J.D. Jackson, {\em Classical Electrodynamics}  $3^{rd}$ Edition, (John Wiley \& Sons, 1999).

\bibitem{fierz} M. Fierz, %``On the theory of magnetically charged particles", 
Helv. Phys. Acta {\bf 17}, 27 (1944).

\bibitem{lipkin} H.J. Lipkin, W.I. Weisberger, and M. Peshkin, %``Magnetic Charge Quantization and Angular Momentum", 
Ann. Phys. {\bf 53}, 203 (1969).

\bibitem{yang} C.N. Yang, %``Monopoles, Fiber Bundles and Gauge Fields", 
Annals New York Academy of Sciences {\bf 294}, 86 - 97 (1977).

\bibitem{singleton-plb} D. Singleton, Phys. Lett. B {\bf 427}, 155 (1998). 

\bibitem{singleton-98} D. Singleton, Am.J.Phys. {\bf 66}, 697 (1998)

\bibitem{thooft} G. 't Hooft, Nucl.Phys.B {\bf 79}, 276 (1974).

\bibitem{polyakov} A. M. Polyakov, JETP Lett. {\bf 20}, 194 (1974).

\bibitem{prasad}  M.K. Prasad and C. M. Sommerfield, Phys.Rev.Lett. {\bf 35}, 760 (1975).

\end{thebibliography}
\end{document}